\documentclass[preprint,showpacs,preprintnumbers,amsmath,amssymb]{revtex4}

\usepackage{graphicx}
\usepackage{dcolumn}
\usepackage{bm}

\begin{document}
\title{Dynamics of the molecular orientation field coupled to ions in two dimensional ferroelectric liquid
crystals}

\author{Robert A. Pelcovits$^{1}$}
\author {Robert B. Meyer$^{2}$}
\author{Jong--Bong Lee$^{2,}$}
\altaffiliation [Present Address: ] {Department of Physics, POSTECH,
Pohang, Kyungbuk 790--784, Republic of Korea.}
\affiliation{$^{1}$Department of Physics, Brown University,
Providence, RI 02912\\ $^{2}$The Martin Fisher School of Physics,
Brandeis University, Waltham, MA 02454 }

\date{\today}
\begin{abstract}

Molecular orientation fluctuations in ferroelectric smectic liquid
crystals produce space charges, due to the divergence of the
spontaneous polarization.  These space charges interact with
mobile ions, so that one must consider the coupled dynamics of the
orientation and ionic degrees of freedom.  Previous theory and
light scattering experiments on thin free-standing films of
ferroelectric liquid crystals have not included this coupling,
possibly invalidating their quantitative conclusions.  We consider
the most important case of very slow ionic dynamics, compared to
rapid orientational fluctuations, and focus on the use of a short
electric field pulse to quench orientational fluctuations.  We
find that the resulting change in scattered light intensity must
include a term due to the quasi-static ionic configuration, which
has previously been ignored.  In addition to developing the
general theory, we present a simple model to demonstrate the role
of this added term.

\end{abstract}

 \maketitle

\section{Introduction}
Quasi-elastic Rayleigh scattering is a powerful method for studying
the molecular orientational fluctuations in liquid
crystals \cite{orsay,galerne}. The added technique of quenching
fluctuations by a short electric field pulse for ferroelectric
Smectic $C^\ast$ (Sm~$C^\ast$) free standing films was first applied for
studying two dimensional phase transitions by Young \textit{et
al.} \cite{Young}. By measuring the time correlation of thermal
fluctuations of the $\mathbf{c}$ director orientation, they determined
the ratio of the bend($K_b$) or splay($K_s$) elastic constant to the
corresponding viscous coefficients($\eta_b,\eta_s$) and the bend
elastic constant to the square of the spontaneous
polarization($P_0$) in the free standing film of a ferroelectric
liquid crystal, that is, $K_b/P_0^2$, $K_s/\eta_s$, and
$K_b/\eta_b$. Refining the light scattering experiment, Rosenblatt
\textit{et al.} \cite{Rosenblatt1,Rosenblatt2} performed absolute
measurements of the elastic constants, spontaneous polarization, and
viscosities, by monitoring the change of intensity of the scattered
light due to quenching of the director fluctuations by a strong
enough external electric field. As shown in Ref.~{\cite{Young}} for
the director aligned along $x$ direction, the intensity of
depolarized light scattered by fluctuations of wave vector $\vec{q}$
is given by
\begin{equation}\label{intensity}
I(\vec{q})\propto \frac{1}{K_s{q_y}^2+K_b{q_x}^2+2\pi
{P_0}^2|q_x|+P_0E},
\end{equation}
where $q_x$ and $q_y$ are the components of $\vec{q}$ in a bend and
a splay mode respectively, and $E$ is an external electric field.
The scattering geometry was arranged such that one wave-vector mode
can be probed at a time, i.e. $q_y=0$ for a bend mode and $q_x=0$
for a splay mode in Eq.~(\ref{intensity}).

However in the light scattering theory for the ferroelectric free
standing liquid crystals the existence of the ionic impurities
dissolved in the materials was ignored. Pindak \textit{et
al.} \cite{Pindak} reported the ionic impurity effect on the
ferroelectric free standing film qualitatively, by analyzing the
change of the 2$\pi$ wall texture due to the external electric
field. The relaxation time of the impurity ion fluctuations in a
thin film is given by \cite{Pindak,Rosenblattphd}
\begin{equation}\label{tau}
\tau=\frac{1}{2\pi \sigma h q+Dq^2},
\end{equation}
where $\sigma$ is conductivity, $h$ is the thickness of a film, $q$
is a wave vector, and $D$ is a diffusion constant. Using the typical
values of liquid crystals, the conduction term in Eq.~(\ref{tau}),
$2\pi \sigma h q$ can be estimated as $2\times 10^{-1}$ sec$^{-1}$
and the diffusion term $Dq^2=9$ sec$^{-1}$. The decay rate of the
fluctuations of the impurity ions is much slower than the
orientation fluctuation time of the director, around 1
ms \cite{Rosenblattphd}.

In a bend mode, the space charge due to divergence of the
spontaneous polarization is screened on very slow time scales by
impurity ions dissolved in ferroelectric liquid crystals. Similarly,
slow variation in the local charge concentration due to ionic
diffusion is rapidly screened by reorientation of the spontaneous
polarization field, which causes reorientation of the director. In
the bulk ferroelectric liquid crystals, Lu \textit{et al.} \cite{Lu0}
reported a very slow relaxation mode compared to a fast decay by
autocorrelation measurements, which is consistent with diffusion
times associated with ionic motions. In subsequent papers, they
examined the coupling between the director distortions and impurity
ion motions in a ferroelectric liquid crystal
theoretically \cite{Lu1} and experimentally \cite{Lu2}.

In this paper, we study the dynamics of the ion-director coupling
in free standing ferroelectric liquid crystal films in the 2D
limit theoretically.  We consider the limit of slow ion dynamics,
and electric field quenching of the rapid orientation
fluctuations, and find that the field induced change in light
scattering intensity must include terms due to the quasi-static
distribution of ions during the short applied field pulse. Without
the added terms, the field quench technique produces invalid
results.  Additional experiments are needed to take account of
this term quantitatively.  They will be reported elsewhere.

This paper is organized as follows: in the next section we present the free energy of the smectic film including Frank elasticity, the coupling of the external electric field to the space charge, the energy associated with fluctuations in the ionic impurity concentration, and the electrostatic energy of the space and ionic charges. In Sec. III we analyze the relaxational dynamics of the coupled director and ionic degrees of freedom, and provide physical insight into the central result of this analysis using a simple spring--based
model. Concluding remarks including comments on the experimental implications of our work are offered in the final section.

\section{Free Energy}
\begin{figure}
\centering
\includegraphics[width=3.0in]{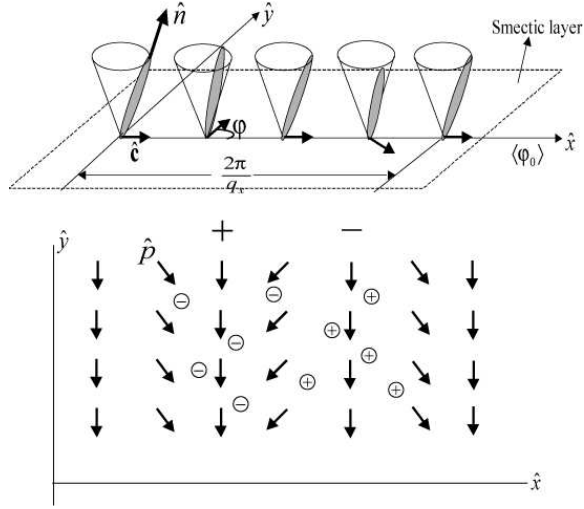}
\caption{Top: Bend mode of the $\mathbf{c}$ director. Bottom: The transverse dipole
$\mathbf{p}$ alignment accompanying the bend mode of the $\mathbf{c}$
director. The arrows denote the direction of the local dipole moment which is perpendicular to the $\mathbf{c}$ director. The impurity ions dissolved in the material are coupled to
the space charge (indicated by the pluses and minuses) \cite{Rosenblattphd}.}\label{fig:fluctuation}
\end{figure}

In the Sm $C^\ast$ phase, there exist both tilt angle and azimuthal
fluctuations \cite{blinc, degenne}. Here we consider temperatures sufficiently below the Sm$C^\ast-A$ transition so that the tilt angle fluctuations are small and we need only study
fluctuations in the azimuthal angle $\varphi$. Therefore, the magnitude of $\mathbf{c}$
director is constant in the film and the molecules fluctuate azimuthally
about $<\varphi_0>$, which is the average azimuthal orientation (see Fig.~\ref{fig:fluctuation}).

We assume that the liquid crystal film lies in the plane $z=0$,
surrounded by vacuum on both sides. Average alignment of the
molecules is achieved by an external electric field small enough
so as not to suppress the thermal fluctuations. In a bend mode of the $\mathbf{c}$
director, the divergence of the transverse dipoles gives rise to space charge. The variation of the space charge due to the
azimuthal fluctuations causes diffusion of the free ions dissolved
in the film. Hence the ionic diffusion is observable in a light
scattering experiment. When the molecular orientation fluctuates
by an angle $\varphi$ from $<\varphi_0>=0$, the $\mathbf{c}$ director
can be expressed by
\begin{equation}\label{eq:cdirector}
\mathbf{c}=(\cos{\varphi(x,y)},\sin{\varphi(x,y)}).
\end{equation}

 The Frank elastic free energy
density of a two--dimensional (2D) Sm $C$ is given by \cite{Young},
\begin{equation}
\nonumber
f_{el}=\frac{1}{2}K_s(\nabla\cdot\mathbf{c})^2+\frac{1}{2}K_b(\nabla\times\mathbf{c})^2,
\end{equation}
which for small fluctuations can be approximated by,
\begin{eqnarray}\label{eq:total}
f_{el}={\frac{1}{2}{K_s\Big(\frac{\partial
\varphi}{\partial y}\Big)^2+\frac{1}{2}K_b(\frac{\partial
\varphi}{\partial x})^2}},
\end{eqnarray}
The splay and bend Frank elastic constants are denoted by $K_s$ and $K_b$, respectively.
The Fourier transform of $f_{el}$ is given by
\begin{equation}\label{f_el(q)}
f_{el}(\mathbf{q}_\perp)=\frac{1}{2}K_sq_{y}^2|\varphi(\mathbf{q}_\perp)|^2+\frac{1}{2}K_bq_{x}^2|\varphi(\mathbf{q}_\perp)|^2.
\end{equation}
where 
\begin{equation}\label{eq:varphiq}
\varphi(\mathbf{q}_\perp)=\int{\varphi(\mathbf{r}_\perp)\exp{(i\mathbf{q}_\perp\cdot
\mathbf{r}_\perp)}}\frac{d^2{r}_\perp}{(2\pi)^2},
\end{equation}
Here we denote the 2D position vector and wavevector by $\mathbf{r}_\perp=(x,y)$ and $\mathbf{q}_\perp=(q_x,q_y)$ respectively. 

 The total electrostatic free energy includes the interaction energy $F_P$ of the polarization $\mathbf{P}$ with the external electric field $\mathbf{E}$ and the electrostatic energy $F_E$ of the space and impurity charges. The Fourier transform of the
 free energy density $f_P$ associated with $F_P$ is given by \cite{Young}:
\begin{equation}\label{f_P(q)}
f_{P}({\mathbf{q}_\perp})=\frac{1}{2}P_0E|\varphi(\mathbf{q}_\perp)|^2,
\end{equation}
where we have again assumed small azimuthal fluctuations and neglected the energy associated with the equilibrium configuration.
Note that this energy density is identical for bend and splay modes.

If we consider $n$ kinds of impurity ions, such that the equilibrium
concentration of the $i^{th}$ type of impurity is $c_i$ and the
local concentration fluctuation is $\delta c_i$, then the free energy density $f_{ion}$ associated with fluctuations of the
impurity ions is given by \cite{Lu1}:
\begin{equation}
f_{ion}(\mathbf{r}_\perp)=\frac{1}{2}k_BT\sum^n_{i=1}{\frac{(\delta
c_i(\mathbf{r}_\perp))^2}{c_i}},
\end{equation}
where $k_B$ is the Boltzmann constant and $T$ is the
temperature. The Fourier transform of this last equation is
\begin{equation}\label{f_i(q)}
f_{ion}(\mathbf{q}_\perp)=\frac{1}{2}k_BT\sum^n_{i=1}{\frac{(\delta
c_i(\mathbf{q}_\perp))^2}{c_i}}.
\end{equation}

The total charge density $\rho(\mathbf{r})$ for an infinitesimally thin film is given by
\begin{equation}
\rho(\mathbf{r})=\sigma(\mathbf{r}_\perp)\delta(z)=[\sigma_{ion}(\mathbf{r}_\perp)+\sigma_P(\mathbf{r}_\perp)]\delta(z),
\end{equation}
where $\sigma_P(\mathbf{r}_\perp)=-\nabla\cdot \mathbf{P}$ is the space charge density due to the
divergence of the spontaneous polarization and the ionic charge density $\sigma_{ion}$ is given by
\begin{equation}
\sigma_{ion}(\mathbf{r}_\perp)=\sum_{i=1}^n e_i\delta c_i(\mathbf{r}_\perp),
\end{equation}
where $e_i$ is the charge of ionic species $i$. Here $\mathbf{r}$ is the full 3D position vector $\mathbf{r}=(x,y,z)$.

The electrostatic free energy of the film (excluding the interaction with the external electric field) is given by:
\begin{equation}
F_e=\frac{1}{2}\int{\rho(\bm{r}) \Phi(\bm{r})d^3{r}},
\end{equation}
where $\Phi$ is the electrostatic potential of the charge density $\rho$ in a dielectric medium with dielectric constant:
\begin{equation}
\label{epsilon} \varepsilon^\prime(z)=1+(\varepsilon-1) a \delta(z).
\end{equation}
Here we assume that the
liquid crystal film has uniform dielectric constant $\varepsilon$ and is surrounded by vacuum on both sides. Mathematically we treat the film as infinitesimally
thin, but introduce the film thickness $a$ in an appropriate
dimensional fashion. 

As shown in Ref.~\cite{lee:051701} the electrostatic free energy can be expressed in Fourier space by

\begin{equation}
F_e=\int f_e(\mathbf{q}_\perp){d^2 q_\perp\over (2\pi)^2},
\end{equation}
where the free energy density $f_e(\mathbf{q}_\perp)$ is given by:
\begin{equation}
f_e(\mathbf{q}_\perp)=\sigma(\mathbf{q}_\perp)\sigma(-\mathbf{q}_\perp){2 \pi
\over 2q_\perp + (\varepsilon -1)a  q_\perp^2},
\end{equation}
which in the long--wavelength limit of experimental relevance ($q_\perp a \ll 1$) simplifies to:
\begin{equation}
\label{fe-q}
f_e(\mathbf{q}_\perp)\approx\sigma(\mathbf{q}_\perp)\sigma(-\mathbf{q}_\perp){
\pi \over q_\perp}.
\end{equation}

The spontaneous polarization $\mathbf{P}$ is given by
$\mathbf{P}=P_{0}[\sin{\varphi}\mathbf {\hat x}-\cos{\varphi}\mathbf{\hat{y}}]$ in the
geometry of Fig.~\ref{fig:fluctuation}. The space charge density
$\sigma_P$ is given for small fluctuations by
\begin{equation}
\sigma_P(x)\approx -P_0\frac{\partial \varphi}{\partial x},
\end{equation}
with Fourier transform:
\begin{equation}
\sigma_P(\mathbf{q}_\perp)\approx iP_0\varphi(\mathbf{q}_\perp)q_x.
\end{equation}
We note that only bend mode fluctuations contribute to the space charge, and we henceforth consider
$\mathbf{q}_\perp=(q,0)$. The
total Fourier--transformed charge density is given by
\begin{equation}\label{eq:charge_q}
\sigma{(q)}=iP_0\varphi(q)q+\sum_{i=1}^n e_i\delta
c_i(\mathbf{q}).
\end{equation}

Substituting Eq.~(\ref{eq:charge_q}) into Eq.~(\ref{fe-q}), we
obtain
\begin{eqnarray}\label{f_e(q)}
f_e(q)&=&\frac{\pi}{q}\Big[iP_0\varphi(q)q+\sum_{i=1}^n
e_i\delta c_i(q)\Big]\Big[-iP_0\varphi^*(q)q+\sum_{j=1}^n
e_j\delta c_j^*(q)\Big]\\
f_e(q)&=&\pi P_0^2q|\varphi(q)|^2+\sum_{i=1}^n\frac{\pi}{q}
e_i^2|\delta c_i(q)|^2+\frac{\pi}{2q} \sum_{i\not= j}e_ie_j
\delta c_i(q) \delta c^\ast_j(q)\nonumber \\&&+i\pi
P_0\sum_{i=1}^ne_i(\varphi (q) \delta c_i^\ast(q) -\varphi^\ast
(q) \delta c_i(q)),
\end{eqnarray}

Using Eqs.~(\ref{f_el(q)}),~(\ref{f_P(q)}),~(\ref{f_i(q)}), and
(\ref{f_e(q)}) the Fourier transformed free energy density is then
given by:
\begin{eqnarray}\label{eq:totalE}
f(q)&=&f_{el}(q)+f_P(q)+f_{ion}(q)+f_e(q)\\
\label{f} &=&\biggl({1\over 2}K_b q^2 +{1\over 2}P_oE+\pi
P_0^2 q\biggr)| \varphi(q)|^2+\sum_{i=1}^n\biggl(\frac{\pi}{q}
e_i^2+\frac{1}{2}\frac{k_BT}{c_i} \biggr)|\delta
c_i(q)|^2\nonumber \\&&+i\pi P_0\sum_{i=1}^ne_i(\varphi (q)
\delta c_i^\ast(q) -\varphi^\ast (q) \delta c_i(q))\nonumber
\\&&+\frac{\pi}{2q} \sum_{i\not= j}e_ie_j \delta c_i(q) \delta
c^\ast_j(q),
\end{eqnarray}

\section{Dynamics}

We now consider the dynamics of the director and ionic fluctuations following the approach of Ref.~\cite{Lu1} where a bulk system was considered. We model the dynamics of the film
with a relaxational equation, assuming a viscosity $\eta$ associated
with bend fluctuations:
\begin{eqnarray}
\nonumber \eta \frac{\partial \varphi(q,t)}{\partial t}
&=&-\frac{\partial f(q,t)}{\partial \varphi_{-q,t}}+ g(t)\\
&=&-(K_b q^2+P_0 E +2\pi P_0^2q)\varphi(q,t)+2\pi
iP_0\sum_{i=1}^ne_i \delta c_i(q,t)+g(t),\nonumber
\\ \label{phieqn}
\end{eqnarray}
where $g(t)$ is a random noise source with zero mean and
autocorrelation function given by:
\begin{equation}
\label{noise}
 <g(t)g(t^\prime)>=2k_BT\eta \delta(t-t^\prime).
\end{equation}

The dynamical equation for the concentration fluctuations is
governed by charge conservation, which in Fourier space reads:
\begin{equation}\label{c-equation}
\frac{\partial \delta c_i(q,t)}{\partial t} =-iq J_{i}(q,t), \ \
i=1,...,n,
\end{equation}
where the current $J_{i}$ is given by
$J_{i}(q,t)=-iqm_ic_i\lbrack\partial f(q,t)/\partial \delta
c_i(-q,t)\rbrack$, and $m_i$ is the mobility of the ion of type
\textit{i}. Thus, Eq.~(\ref{c-equation}) can be written as:
\begin{eqnarray}
\frac{\partial \delta c_i(q,t)}{\partial t}
&=&-m_ic_iq^2\frac{\partial f(q,t)}
{\partial\delta c_i(-q,t)}, \ \ i=1,...,n,\\
&=&-m_ik_BTq^2\delta c_i(q,t)-2\pi m_ic_ie_iq \sigma(q,t), \ \
i=1,...,n, \label{ceqn}
\end{eqnarray}
where we have used Eq.~(\ref{eq:charge_q}).

We solve Eqs.~(\ref{phieqn}) and (\ref{ceqn}) by Laplace
transforming in time. To simplify the calculation we assume that the
ionic mobility $m_i$ is independent of the ion type \textit{i}; we
denote this common value by \textit{m}. We introduce the Laplace
transforms of $\varphi, \delta c_i,g(s)$, and $\sigma$ as follows:
\begin{eqnarray}
\varphi(q,s)&=&\int^\infty_0 dte^{-st}\varphi(q,t) \label{l1}\\
\delta c(q,s)&=&\int^\infty_0dt e^{-st}\delta c(q,t) \label{l2}\\
\sigma(q,s)&=&\int^\infty_0 dte^{-st}\sigma (q,t) \label{l3}\\
g(s)&=&\int^\infty_0dt e^{-st}g(t)\label{l4}
\end{eqnarray}

Using Eqs.~(\ref{l1})--(\ref{l4}), we Laplace transform
Eqs.~(\ref{phieqn}) and (\ref{ceqn}) and find:
\begin{eqnarray}
\eta(s\varphi(q,s)-\varphi_o(q))&=&-(K_b q^2+P_0 E +2\pi
P_0^2q)\varphi(q,s)\nonumber \\
&&+2\pi iP_0\sum_{i=1}^ne_i \delta c_i(q,s)+g(s)\label{phil}\\
s\delta c_i(q,s)-\delta c_{io}(q)&=&-m_ik_BTq^2\delta
c_i(q,s)-2\pi m_ic_ie_iq \sigma(q,s),\nonumber\\
&&i=1,...,n,\label{cl}
\end{eqnarray}
where, $\varphi_o(q)\equiv \varphi(q,t=0)$, and $\delta
c_{io}(q)\equiv \delta c_i(q,t=0)$.

We eliminate $\delta c_i(q,s)$ from Eq.~(\ref{phil}) using the
definition of $\sigma(q,s)$, Eq.~(\ref{eq:charge_q}):
\begin{equation}
\eta(s\varphi(q,s)-\varphi_o(q))=-(K_b q^2+P_0 E)\varphi(q,s)+2\pi
iP_0\sigma(q,s)+g(s)\label{phil2},
\end{equation}
and eliminate $\delta c_i(q,s)$ from Eq.~(\ref{cl}) by first
multiplying the latter equation by $e_i$ and then summing over
\textit{i}. Using Eq.~(\ref{eq:charge_q}) we obtain:
\begin{equation}\label{rhos}
(s+m
k_BTq^2)(\sigma(q,s)-iP_0q\varphi(q,s))=\sigma_o(q)-iP_0\varphi_o(q)-2\pi
mq\sigma (q,s)\sum_{i=1}^ne^2_ic_i,
\end{equation}
where $\sigma_o(q)\equiv \sigma(q,t=0)$.

Finally, eliminating $\sigma(q,s)$ from Eqs.~(\ref{phil2}) and
(\ref{rhos}) we obtain the following solution for $\varphi(q,s)$:
\begin{equation}
\label{phis} \varphi(q,s)={g(s)+\eta\varphi_o(q)+2\pi i
P_0{\sigma_o(q)-iP_0q\varphi_o(q)\over s+mk_BTq^2+2\pi
mq\sum_{i}e^2_ic_i}\over \eta s+K_bq^2+P_0E+{2\pi
P_0^2\over\frac{2\pi m}{s+mk_BTq^2}\sum_{i}e^2_ic_i+\frac{1}{q}}}
\end{equation}

This expression for $\varphi(q,s)$ is of the form:
\begin{equation}
\varphi(q,s)=A(s)g(s)+B(s),
\end{equation}
where:
\begin{eqnarray}
A(s)&=&1 \over \eta s+K_bq^2+P_0E+{2\pi P_0^2\over\frac{2\pi m}
{s+mk_BTq^2}\sum_{i}e^2_ic_i+\frac{1}{q}}\\
B(s)&=& {\eta\varphi_o(q)+2\pi i
P_0{\sigma_o(q)-iP_0q\varphi_o(q)\over s+mk_BTq^2+2\pi
mq\sum_{i}e^2_ic_i}\over \eta s+K_bq^2+P_0E+{2\pi
P_0^2\over\frac{2\pi
m}{s+mk_BTq^2}\sum_{i}e^2_ic_i+\frac{1}{q}}}\label{bs}.
\end{eqnarray}

The convolution theorem for Laplace transforms then yields the
following solution for $\varphi$ as a function of time:
\begin{equation}
\label{laplacesoln} \varphi(q,t)={\cal
L}^{-1}[B(s)]+\int^t_oA(t-t^\prime)g(t^\prime)dt^\prime,
\end{equation}
where the operator ${\cal L}^{-1}$ is the inverse Laplace transform,
and $A(t)={\cal L}^{-1}[A(s)]$.

We evaluate the inverse Laplace transforms appearing in
Eq.~(\ref{laplacesoln}) using the Bromwich integral:
\begin{equation}
\label{bromwich} {\cal L}^{-1}[A(s)]=\sum\mbox{residues of the poles
of}\ A(s)e^{st}.
\end{equation}

The functions $A(s)$ and $B(s)$ have identical simple poles at
$s=s_1,s_2$:
\begin{equation}
s_{1,2}={-\alpha\pm\sqrt{\alpha^2-4\eta\beta}\over 2 \eta},
\end{equation}
where
\begin{eqnarray}
\alpha&=&2\pi P_0^2q+K_bq^2+P_0E+2\pi\eta mq\sum_ie_i^2c_i+\eta m k_BTq^2,\\
\beta&=&(K_bq^2+P_0E)(mk_BTq^2+2\pi m q\sum_ie_i^2c_i)+2\pi
P_0^2mk_BTq^3
\end{eqnarray}

It is instructive to examine some limiting cases of these poles as
was done in Ref.  \cite{Lu1}.
\paragraph{Static ions: $m=0$.}
In this case the locations of the poles are given by:
\begin{equation}
\label{staticion} s_1=-\eta^{-1}(K_bq^2+P_0E+2\pi P_0^2q),\ \ s_2=0,
\end{equation}
i.e., $s_1$ describes the director relaxation rate in the absence of
ions, while $s_2$ corresponds to the infinite relaxation time of the
static ions.
\paragraph{Static director: $K_b=P_0=0$.}
Here the poles are given by:
\begin {equation}
s_1=0,\ \ s_2=2\pi mq\sum_ie^2_ic_i+mk_BTq^2,
\end{equation}
where $s_2$ is the relaxation rate of the ions, and $s_1$ describes
the static director. These results agree with the corresponding
results found in Ref.  \cite{Lu1} for the bulk ferroelectric liquid
crystal.

Returning to Eq.~(\ref{laplacesoln}), we evaluate
$<|\varphi(q,t)|^2>$, a quantity proportional to the scattered light
intensity. The brackets refer to an average over the Boltzmann
ensemble of $\phi_o$ and $\sigma_o$ (which appear in $B(s)$), and
the random noise source $g(t)$, whose variance is given by Eq.
(\ref{noise}). We find:
\begin{equation}
\label{phisq} <|\varphi(q,t)|^2>=<|B(q,t)|^2>+2\eta k_BT\int^t_o
A^2(t-t^\prime) dt^\prime.
\end{equation}

While the averages and integral in Eq.~(\ref{phisq}) can in
principle be evaluated for arbitrary values of the ionic mobility,
bend elastic constant and polarization, the expressions obtained are
rather complicated, so we consider instead the experimentally
relevant case where  the ionic mobility $m\rightarrow 0$ and assume
that the electric field is switched on at  $t=0^+$. Using
Eqs.~(\ref{eq:charge_q}), (\ref{phis}), (\ref{bromwich}) and
(\ref{staticion}), we find from Eq.~(\ref{phisq}):
\begin{eqnarray}
\label{phisqsimple} \lim_{t\to\infty}<|\varphi(q,t)|^2>&=&{(2\pi
P_0)^2 \over (K_bq^2+P_0E+2\pi P_0^2q)^2}<|\sum_ie_i\delta c_i|^2>_o
\nonumber \\&&+{k_BT\over K_bq^2+P_0E+2\pi P_0^2q},
\end{eqnarray}
where the thermal average is over the Boltzmann ensemble at $t=0$
when $E=0$.  Using Eq.~(\ref{f}) we
find:
\begin{equation}
\label{cfluc}
<|\sum_ie_i\delta c_i|^2>_o=k_BT {K_bq+2\pi P_0^2 \over
4\pi\lambda_{2D}(K_bq+2\pi P_0^2)+2\pi K_b}
\end{equation}
where $\lambda_{2D}$ is the Debye screening length in 2D defined by
\begin{equation}
\lambda_{2D}\equiv\frac{k_BT}{4\pi\sum_{i}c_ie_{i}^2}.
\end{equation}

The expression Eq.~(\ref{phisqsimple}) for $<|\varphi(q,t)|^2>$ can
be given a simple physical interpretation. In the limit of static
ions where $m=0$, the director angle $\varphi$ has a mean value
given by:
\begin{equation}
\label{varmean} <\varphi(q,t)>=<B(t)>= {2\pi i P_0 \sum_ie_i\delta
c_i\over K_bq^2+P_0E+2\pi P_0^2q}
\end{equation}
using Eqs.~(\ref{bs}) and (\ref{laplacesoln}), and recalling that
the noise source $g(t)$ has zero mean. More simply, this result can
be obtained by averaging Eq.~(\ref{phieqn}) over the noise and
noting that $<{\partial \varphi(q,t)\over \partial t}>=0$.

We now write $\varphi$ as:
\begin{equation}
\varphi(q,t)=<\varphi(q,t)>+(\varphi(q,t)-<\varphi(q,t)>),
\end{equation}
and note that the fluctuation of $\varphi$ about its mean value,
Eq.~(\ref{varmean}), has a mean-squared average:
\begin{equation}
\label{varfluc} <|\varphi(q,t)-<\varphi(q,t)>|^2>={k_BT\over
K_bq^2+P_0E+2\pi P_0^2q},
\end{equation}
as can be seen using Eq.~(\ref{f}).

Then, it can be readily seen that the mean--squared average of
$\varphi$,
\begin{equation}
\label{varphi}
<|\varphi(q,t)|^2>=|<\varphi(q,t)>|^2+<|\varphi(q,t)-<\varphi(q,t)>|^2>,
\end{equation}
yields Eq.~(\ref{phisqsimple}) in the long--time limit. Note that
because the ions are static, the application of the electric field
at $t=0$ has no effect on the value of $<|\sum_ie_i\delta c_i|^2>_o$
which enters the first term on the right--hand side of
Eq.~(\ref{varphi}).

\begin{figure}
\centering
\includegraphics[width=2.5in]{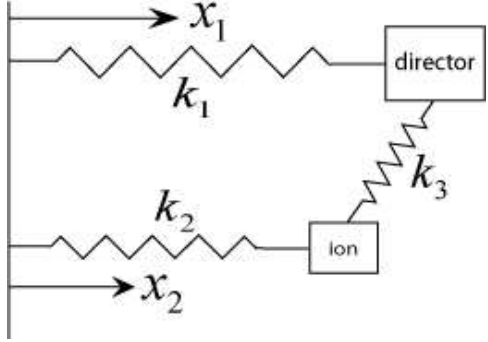}
\caption{A toy model for the coupling between director and ionic degrees of freedom.}\label{fig:toy}
\end{figure}
Additional physical insight into Eq.~(\ref{phisqsimple}) can be obtained by considering the toy model shown in
Fig.~\ref{fig:toy}. We represent the director mode by a single variable $x_1$ and the ionic displacement mode by the single variable $x_2$. The spring constants $k_1$ and $k_2$ represent the corresponding restoring forces for these modes. A third spring constant $k_3$ represents the coupling of the director and ion fluctuations. The energy of director fluctuations in this model in the absence of an external electric field is given by,
\begin{eqnarray}\label{springE}
\nonumber F&=&\frac{1}{2}k_1x_1^2+\frac{1}{2}k_3(x_2-x_1)^2\\
&=&\frac{1}{2}(k_1+k_3)x_1^2-k_3x_1x_2+\frac{1}{2}k_3x_2^2.
\end{eqnarray}
The equilibrium value of $x_1$ which we denote by $\bar{x}_1$ is given by solving $\partial F/\partial x_1=0$ with the result:
\begin{equation}\label{x1}
\bar{x}_1=\frac{k_3}{k_1+k_3}x_2.
\end{equation}
Defining the variation $y=x_1-\bar{x}_1$ of $x_1$ about its equilibrium position $\bar{x_1}$, the
spring free energy in Eq.~(\ref{springE}) can be rewritten as
\begin{equation}
F=\frac{1}{2}(k_1+k_3)y^2+\textrm{constant}.
\end{equation}
Using the equipartition theorem the thermal average of the square of $y$ is given by
\begin{equation}
<y^2>=\frac{k_BT}{k_1+k_3}.
\end{equation}
and the corresponding quantity for $x_1$ is given by
\begin{eqnarray}
\nonumber <x_1^2>&=&<\bar{x}_1^2+y>\\
&=&<\bar{x}_1^2>+ 2<\bar{x}_1y>+<y^2>\\
\label{x1noE}&=&\frac{k_3^2}{(k_1+k_3)^2}<x_2^2>+\frac{k_BT}{k_1+k_3},
\end{eqnarray}
where $<\bar{x}_1y>=0$ because $x_1$ and $y$ are statistically independent in our model. This equation is analogous to Eq.~(\ref{varphi}) above. Now, imagine a sudden application of the external electric field $E$ which leads to the replacement of the spring constant $k_1$ by $k_1+E$. If we assume that the free ions have a very long decay time then
$x_2$ can be considered a constant during the electric field pulse. Hence with the application of the electric field, $<x_1^2>$ is given by
\begin{equation}\label{eq:toymodel}
<x_1^2(E\neq
0)>=\frac{k_3^2}{(k_1+E+k_3)^2}<x_2^2>+\frac{k_BT}{k_1+E+k_3}.
\end{equation}
This expression is analogous to our central result Eq.~(\ref{phisqsimple}) above. The second term on the right--hand side of Eq.~(\ref{eq:toymodel}) corresponds to $<y^2>$, the fluctuation of $x_1$, the director mode, about its mean value $\bar{x}_1$, while the first term corresponds to $<\bar{x}_1^2>$, which in turn depends on the ionic degree of freedom $x_2$, as indicated in Eq.~(\ref{x1}).

\section{Conclusions}
In this paper we have considered the coupled dynamics of the orientational and ionic degrees of freedom in thin freely--suspended smectic liquid crystal films. Our central result shown in Eqs.~(\ref{phisqsimple})--(\ref{cfluc}) describes the fluctuations in the azimuthal angle of the $\mathbf {c}$ director, which is proportional to the scattered light intensity. As illustrated in our toy model at the end of the last section, the fluctuations can be understood as arising from two contributions: the director fluctuations measured relative to their mean value (the second term on the right--hand sides of Eqs.~(\ref{phisqsimple}) and (\ref{eq:toymodel})) and the change in this mean value due to coupling to the ions (the first term on the right--hand side of Eqs.~(\ref{phisqsimple}) and (\ref{eq:toymodel})). Previous theoretical and experimental work on smectic films which ignored the existence of ionic impurities thus ignored the second contribution to the light scattering intensity.  While we have shown that in principle the ionic impurities will modify the light scattering intensity the effect might in practice be negligible. Upon using Eq.~(\ref{cfluc}) the ionic contribution to the director fluctuations (the first term on the right--hand side of Eq.~{(\ref{phisqsimple})}) is given by: 
\begin{equation}\label{coupling}
\frac{(2\pi P_0)^2}{(K_bq^2+P_0E+2\pi P_0^2q)^2}\bigg[\frac{k_BT
(K_bq+2\pi P_0^2)}{4\pi \lambda_{2D} (K_bq+2\pi P_0^2)+2\pi
K_b}\bigg].
\end{equation}
If the screening length $\lambda_{2D}$ is shorter than a few microns
 at wavevectors in the range $2000 \sim 5000 cm^{-1}$, the ionic contribution is not small compared to the second term in Eq.~(\ref{phisqsimple}).
However, if $\lambda_{2D}\gtrsim 10 \mu$m, then the ionic contribution, Eq.~(\ref{coupling}), contributes less than 10$\%$
 to the expression Eq.~(\ref{phisqsimple}) for the director fluctuations.  Therefore for any
particular experiment one must evaluate the relative importance of
the ionic contribution shown in Eq.~(\ref{coupling}).

\section*{Acknowledgments}

R.A.P. was supported in part by the NSF under Grant No.
DMR-0131573. This research was supported at Brandeis University by
NSF Grant No. DMR- 0322530.

\end{document}